\documentclass[preprint,showpacs,preprintnumbers,amsmath,amssymb]{revtex4-1}
\usepackage{graphicx,textcomp,caption,subcaption}
\usepackage{dcolumn}
\usepackage{bm}
\usepackage{amsmath}

\begin{document}

\title{Resonant states of H$_3^-$ molecule and its isotopologues D$_2$H$^-$ and H$_2$D$^-$.}
\author{M. Ayouz$^{1}$, O. Dulieu$^{2}$, and J. Robert$^{2}$}
\affiliation{$^{1}$Laboratoire de G\'enie des Proc\'ed\'es et Mat\'eriaux, Ecole Centrale de Paris, B\^at. Dumas, 92295 Ch\^atenay-Malabry Cedex, France\\ $^{2}$Laboratoire Aim\'e Cotton, CNRS/Univ Paris-Sud/ENS Cachan, B\^at. 505, Campus d'Orsay, 91405 Orsay Cedex, France}

\begin{abstract}
Using the ground potential energy surface[M. Ayouz \textit{et\, al}. J. Chem Phys \textbf{132} 194309 (2010)] of the H$_3^-$ molecule, we have determined the energies and widths of the complex resonant levels of H$_3^-$ located up to 4000~cm$^{-1}$ above the dissociation limit H$^-$+H$_2(v_d=0,j_d=0)$. Bound and resonant levels of the H$_2$D$^-$ and D$_2$H$^-$ isotopologues have been also characterized within the same energy range. The method combines the hyperspherical adiabatic approach, slow variable discretization method, and complex absorbing potential. These results represent the first step for modeling the dynamics of the associated diatom-negative ion collision at low energy involving rotational quenching of the diatom and reactive nucleus exchange via weak tunneling effect through the potential barrier of the potential energy surface.
\end{abstract}

\pacs{34.80.Ht 34.80.Lx}

\maketitle

\section{Introduction}

About 92\% of the atoms in the universe are hydrogen atoms, so that the chemistry of the interstellar medium (ISM) is dominated by its amount under various molecular forms and charge states. It is thus enlightening to briefly review a few of them. The H$_2$ molecule is the most abundant component of the interstellar gas. It is constantly ionized by cosmic rays to produce strongly bound H$_2^+$ ions which rapidly reacts with H$_2$ to produce H$_3^+$ \cite{oka2006}. While much smaller than the H$_2$ one, the steady-state concentration of H$_3^+$ ion is the largest of all molecular species in the ISM. The H$_2$ molecule has a proton affinity of about 4.5~eV \cite{mccall2004} smaller than the one of most molecules, so that H$_3^+$ represents a very efficient proton donor inducing then a rich chemistry. The simplest negative ion H$^-$ is thought to be present in noticeable amounts in the ISM, but it has not been detected yet despite huge observational efforts \cite{ross2008}. The H$_2^-$ anion is known to be unstable, and exists only as a resonance induced by high rotational states in H$_2$ \cite{golser2005}. The neutral H$_3$ molecule has no bound level as it exhibits a purely repulsive potential energy surface in its ground state \cite{wu1999}.

While the H$_3^+$ ion is often presented as the simplest triatomic molecule, it is interesting to recall that the H$_3^-$ anion exists as well, while its detailed structure has not been characterized yet by spectroscopy. This ion has been observed in a discharge plasma experiment \cite{wang2003} and in sputtering experiments \cite{gnaser2006} using mass spectrometry, together with various isotopic compounds.  In a previous study \cite{ayouz2010} (hereafter referred to as paper I) we theoretically investigated the structure of the electronic ground state H$_3^-$ anion, including its PES and its permanent dipole moment surface (PDMS). Following previous studies of the same type \cite{starck1993}, we confirmed that this ion is bound due to long-range polarization forces induced by the interaction of the quadrupole moment of H$_2$ with the charge of H$^-$. For this reason it is often referred to as a van der Waals molecule. We determined the energy of the rovibrational bound levels of its electronic ground state, the deepest one being bound by about 71~cm$^{-1}$ with respect to the H$^-$+H$_2$($v_d=0, j_d=0$) dissociation limit. In a further study \cite{ayouz2011} (hereafter referred to as paper II) we explored the formation of the H$_3^-$ anion in the ISM via the radiative association (RA) of H$_2$ and H$^-$. If probed, the presence of H$_3^-$  would then be a proof of the still unresolved issue of the existence of H$^-$ itself in the ISM. Unfortunately, the computed rate for the RA reaction H$_2$ + H$^- \rightarrow$ H$_3^-$ + (photon) has been found too weak to expect a significant signal from the absorption spectrum of  H$_3^-$.

Besides, the interest for the study of low-energy chemical reaction between negative ions and neutral systems is currently increasing \cite{mikosch2010} with the development of multipole radio-frequency traps for ions \cite{wester2009} combined with buffer-gas-cooling and laser-cooling techniques \cite{willitsch2008}. In this respect, the elastic, inelastic, and reactive collisions between H$^-$ and H$_2$ and between their various isotopologues (D$^-$, HD, D$_2$) represent prototype processes for such fundamental systems, which have been extensively studied both experimentally (see for instance Refs.\cite{haufler1997}, \cite{mikosch2010}, paper I, and references therein) and theoretically (see for instance Refs.\cite{gianturco1995,takayanagi2000,jaquet2001,giri2007}, paper I, and references therein). Heteronuclear systems like H$_2$+D$^-$ and HD+H$^-$ are particularly relevant as they allow the detection of reactive processes involving the exchange of one H and one D atom. To our knowledge, besides our study of paper II, there is only one other theoretical work devoted to the dynamics of the bound H$_2$D$^-$ van der Waals system, namely its prereaction induced by a photonic excitation \cite{takayanagi2000} of resonances lying within the electronic ground state. Multiple rf traps would be the perfect tool to study the formation of such triatomic anions through radiative processes and their level structure: indeed the temperature of the trap can nowadays be controlled and varied with a good precision while the trapped ions can be kept inside the trap for hours \cite{wester2009,wester_pc}, allowing for a long observation ideal for low rate processes like radiative association.

To guide such prospected experimental studies, it is highly desirable to theoretically investigate the level structure of this family of triatomic anions, taking advantage of the new ground state potential energy surface determined in paper I. The present paper is devoted to the characterization of the stable levels and the prereacting resonances of H$_3^-$ and its isotopologues lying within their electronic ground state. In Section \ref{sec:method}, we briefly recall the main steps of the bound state calculations based on the hyperspherical adiabatic approach, and on the resolution of the Schr\"odinger equation with the Slow Variable Discretization (SVD) method. Energies and (non-radiative) widths of resonant states are obtained through the use of a complex absorbing potential (CAP). As a follow-up of papers I and II, the energy and widths of the prereacting resonances of H$_3^-$ are presented in Section \ref{sec:h3-} up to the H$_2$($v_d=1$,$j_d=1$) collisional threshold. We display in Section \ref{sec:hetero} the hyperspherical adiabatic potential curves and the calculated energies of the bound levels of the H$_2$D$^-$ and D$_2$H$^-$ systems, and the energies and widths of their prereacting resonances. Atomic units for distances (1~a.u.=$a_0$=0.0529177~nm) and energies (1~a.u.=2 $R_{\infty}$=219474.63137~cm$^{-1}$ will be used throughout the paper except otherwise stated.

\section{The SVD formulation of the Schr\"odinger equation}
\label{sec:method}

The system of Smith-Whitten hyperspherical collective coordinates \cite{johnson1980} is a well-known convenient starting point to write the Schr\"odinger equation for a triatomic system composed of three particles with masses $m_i$ ($i=$1,2,3) with total mass $M$ and reduced mass $\mu$:
\begin{equation}
M=\sum_{i=1}^{3}m_i; \mu=\sqrt{\frac{\Pi_{i=1}^{3}m_i}{M}},
\label{eq:masses}
\end{equation}
The hyper-radius $\rho$ and the hyperangles $\theta$ and $\phi$ are related to the three relative internuclear distances $r_i$ ($i=$1,2,3) according to
\begin{equation}
r_i=\frac{\rho d_i}{\sqrt{2}}\sqrt{1+\sin\theta\sin(\phi+\epsilon_i)}
\label{eq:r_i}
\end{equation}
where
\begin{equation}
d_i=\sqrt{\frac{m_i}{\mu}\left(1-\frac{m_i}{M}\right)}
\end{equation}
\begin{eqnarray}
\epsilon_1&=&2\arctan\left(\frac{m_2}{\mu}\right)\\
\epsilon_2&=&-2\arctan\left(\frac{m_1}{\mu}\right)	
\end{eqnarray}
and $\epsilon_3=0$. The main steps of our approach are discussed in detail in several publications \cite{kokoouline2006,blandon2007,ayouz2010} and are briefly recalled below. It relies on the adiabatic separation between hyper-radius and hyperangles combined to the slow variable discretization (SVD) method \cite{tolstikhin1996,kokoouline2006,blandon2007} which allows for an easy treatment of non-adiabatic couplings between hyperspherical adiabatic channels. Assuming a total angular momentum $J=0$, the eigenfunctions $\Psi=\rho^{-5/2}\psi$ with energy $\mathcal{E}$ of the three-particle system mutually interacting through a potential $V$ depending only on the three internuclear distances are obtained by solving the Schr\"odinger equation expressed  in the center-of-mass frame \cite{johnson1980}
\begin{eqnarray}
\left(-\frac{1}{2\mu}\frac{\partial^2}{\partial \rho^2}+\frac{\Lambda^2_0+\frac{15}{4}}{2\mu\rho^2}+V\right)\psi=\mathcal{E}\psi	
\label{eq:Schrodinger_Equation}
\end{eqnarray}
The operator $\Lambda_0$ above is the grand angular momentum\cite{smith1960b,johnson1980} depending only on $\theta$ and $\phi$ for $J=0$. This equation is solved in two steps. Assuming a discrete variable representation (DVR) of the hyperspherical radius on $N_{\rho}$ points ($j=1,...,N_{\rho}$) the following 2D Schr\"odinger equation is solved at every $\rho_j$
\begin{eqnarray}
\left(\frac{\Lambda^2_0+\frac{15}{4}}{2\mu\rho_j^2}+V\right)\varphi_{a,j}(\theta,\phi)=U_a(\rho_j)\varphi_{a,j}(\theta,\phi)
\label{eq:ad-states}
\end{eqnarray}
The function $\varphi_{a,j}(\theta,\phi)$ is the hyperangular wave function at $\rho_j$ for $a$-th eigenvalue $U_a(\rho_j)$, and is expanded over a set of B-splines $h_{\kappa}(\phi)$ and $g_{\lambda}(\theta)$ \cite{boehm2002,esry1997,ayouz2010b}
\begin{equation}
\varphi_{a,j}(\theta,\phi)=\sum_{\kappa,\lambda}C_{\kappa,\lambda}h_{\kappa}(\phi)g_{\lambda}(\theta)
\end{equation}
The members of the collection of $U_a(\rho_j)$ values (with $1 \leq a \leq N_c$, $N_c$ being the number of channels included in the calculation) are called adiabatic hyperspherical potentials. The total vibrational eigenfunction $\psi(\rho,\theta,\phi)$ is written as an expansion over the basis functions $y_{a,j}(\rho,\theta,\phi)$ with unknown coefficients $c_{a,j}(\rho)$:
\begin{equation}
 \psi(\rho,\theta,\phi)=\sum_{a=1}^{N_c} \sum_{j=1}^{N_{\rho}} y_{a,j}(\rho,\theta,\phi)c_{a,j}(\rho)\,
\label{eq:expansion}
\end{equation}
The basis functions $y_{a,j}(\rho,\theta,\phi)$ are constructed as products of $\varphi_{a,j}(\theta,\phi)$ with plane wave functions $\pi_j(\rho)$ representing a set of orthogonal and normalized basis functions localized at DVR grid points $\rho_j$:
\begin{equation}
y_{a,j}(\rho,\theta,\phi)= \pi_j(\rho)\varphi_{a,j}(\theta,\phi)\,
\end{equation}
Inserting the expansion of Eq.\ref{eq:expansion} in Eq.\ref{eq:Schrodinger_Equation} leads to a set of coupled differential equations for the coefficients $c_{a,j}$ associated to the eigenvalue $\mathcal{E}$
\begin{eqnarray}
\sum_{a'j'}\left[\langle\pi_{j'}\arrowvert -\frac{1}{2\mu}\frac{\partial^2}{\partial\rho^2}\arrowvert \pi_j\rangle{O}_{a'j',aj}+U_a(\rho)\delta_{a'a}\delta_{j'j} \right]c_{j'a'}(\rho)=
\mathcal{E} \sum_{a'}O_{a'j,aj}c_{ja'} \,,
\label{eq:gen_eigen_value}
\end{eqnarray}
where the overlap matrix elements $O_{a'j,aj}$ are defined as
\begin{eqnarray}
{O}_{a'j',aj}=\langle\varphi_{a',j'}(\theta,\phi)\arrowvert\varphi_{a,j}(\theta,\phi)\rangle\,.
\label{eq:overlap}
\end{eqnarray}
where the brackets denote integration over hyperangles $\theta,\phi$.

The main advantage of the SVD method \cite{kokoouline2006,blandon2007} is that these matrix elements account for the non-adiabatic couplings between the hyperspherical channels through the matrix elements of Eq. \ref{eq:overlap}, avoiding to consider the derivatives of adiabatic states with respect to $\rho$ and therefore greatly simplifying the numerical solution of the equation. In fact, the SVD method is similar in spirit to the diabatic-by-sector method by Launay and co-workers (see for instance Ref.\cite{soldan2002}). In both methods, at each hyper-radius $\rho$, an adiabatic $\varphi_{a,j}(\theta,\phi)$ basis is calculated. In the diabatic-by-sector method, a Numerov-type propagation procedure is implemented over each $\rho$ sector using this locally-adapted adiabatic basis. In the SVD approach, the locally-adapted adiabatic basis is directly used to construct the global Hamiltonian matrix, which is then diagonalized, resulting into a relatively small-sized Hamiltonian matrix.

In order to adapt the DVR of $\rho$ to the local oscillation frequency of the hyperradial coefficients $c_{ja}(\rho)$ and to minimize the number of DVR grid points, we use a mapping procedure $\rho=\rho(x)$ of the hyperradius coordinate by a new coordinate $x$ \cite{kokoouline1999}. The grid steps $\Delta\rho$ and $\Delta x$ in $\rho$ and $x$ are linked by the jacobian $\jmath(x)$ of this change of variable
\begin{equation}
\Delta\rho=\jmath(x)\Delta x\, \qquad \text{where}\qquad \jmath(x)=\frac{d\rho}{dx}
\end{equation}
Equation \ref{eq:gen_eigen_value} is then rewritten as
\begin{eqnarray}
\sum_{a'j'}\left[\langle\pi_{j'}\arrowvert -\frac{1}{4\mu}\left(\frac{1}{\jmath^2}\frac{d^2}{dx^2}+\frac{d^2}{dx^2}\frac{1}{\jmath^2}\right)\arrowvert \pi_j\rangle{O}_{a'j',aj}+\tilde{U}_a(\rho(x))\delta_{a'a}\delta_{j'j} \right]\tilde{c}_{j'a'}=
\mathcal{E} \sum_{a'}O_{a'j,aj}\tilde{c}_{ja'} \,,
\label{eq:gen_eigen_value_mapping}
\end{eqnarray}
where the transformed adiabatic potentials $\tilde{U}_a(\rho(x))$ depend on the derivatives $\jmath'$ and $\jmath''$ with respect to $x$
\begin{eqnarray}
\tilde{U}_a(\rho(x))=U_a(\rho(x))+\frac{1}{4\mu}\left(\frac{7}{2}\frac{(\jmath')^2}{\jmath^4}-\frac{\jmath''}{\jmath^3}\right)
\end{eqnarray}
The coefficients $\tilde{c}_{j'a'}=\sqrt{\jmath} c_{j'a'}$ correspond to the definition of a scaled wave function $\tilde{\psi}(\rho,\theta,\phi)=\sqrt{\jmath}\psi(\rho,\theta,\phi)$ normalized to 1 in the $x$ coordinate. In practice, the mapping function $\rho(x)$ is estimated from the local de Broglie wavelength, which is given from the local kinetic energy $E_{kin}(\rho)$ so that
\begin{eqnarray}
\Delta\rho=\beta\frac{\pi}{2\mu E_{kin}(\rho)}	
\end{eqnarray}
where the parameter $\beta\leq 1$ allows for checking convergence of the results against the size of the grid.

Eq.\ref{eq:gen_eigen_value_mapping}  is a generalized eigenvalue problem of dimension $(N_{\rho}\times N_c)\times (N_{\rho}\times N_c)$ with eigenvalues $\mathcal{E}$ eigenvectors $\tilde{c}$. It is solved using standard numerical procedures taken from LAPACK and ARPACK libraries \cite{netlib}.

Resonant states and their non-radiative lifetime are obtained using a complex absorbing potential (CAP) which simulates an infinite hyperradial grid (see Refs.\cite{vibok1992,riss1996}). The CAP is placed at the end of the hyperradial grid and absorbs the outgoing dissociation flux of decaying resonant states. The same CAP is used for adiabatic hyperspherical potentials \textit{i.e.} $\tilde{U}_a(\rho(x))\rightarrow \tilde{U}_a(\rho(x))-iV_{CAP}(\rho(x))$. The resulting Hamiltonian is now non-Hermitian and its complex eigenenergies are expressed as $\mathcal{E}=E_r-i \Upsilon/2$ where $E_r$ is the energy position of the resonance and $\Upsilon$ is the resonance width. Following Ref.\cite{vibok1992} and the study of Blandon \textit{et al.} \cite{blandon2007} about resonances in three-body systems, we used the quadratic form
\begin{equation}
V_{CAP}(\bar{\rho})=\frac{3}{2}A_2\bar{\rho}^2
\end{equation}
with
\begin{equation}
\bar{\rho}=\frac{\rho-(\rho_{max}-L)}{L}
\end{equation}
where $L$ is the damping length and $A_2$ the magnitude of the CAP, and $\rho_{max}$ the upper bound of the hyperradial grid.

The parameters of the CAP ($L=12$~a.u., $A_2=0.0046$~a.u.) are adjusted in order to minimize the sum of the reflection and transmission coefficients of the wave functions according to the general recommendations of Ref.\cite{vibok1992}. We should recall that solving Eq.\ref{eq:gen_eigen_value_mapping} delivers complex energies related either to resonances, or to discretized continuum states related to energetically-open dissociation channels. Resonances are identified by the convergence and of their energy and width with respect to changes of $N_{\rho}$, $N_c$, $N_{\theta}$, $N_{\phi}$ $L$, and $A_2$. We found that $N_{\rho}=180$ and $N_{\theta}=60$ were sufficient to converge the adiabatic hyperspherical potentials $U_a(\rho)$ for all ions. $N_{\phi}=80$ was used for H$_3^-$ while $N_{\phi}=140$ was needed for H$_2$D$^-$ and D$_2$H$^-$ due to the strong localization of the hyperangular wave functions. Finally the convergence of $E_r$ values was also checked especially for high-lying resonant states (close to 4000~cm$^{-1}$ above $E^d_{00}$) leading to $N_c=10$ for H$_3^-$ and $N_c=20$ for H$_2$D$^-$ and D$_2$H$^-$ (due to the important number of dissociation channels in the latter case).

\section{The prereacting resonances of the H$_3^-$ ground state}
\label{sec:h3-}

In the remaining part of the paper we use the ground state potential energy surface (PES) determined in paper I with the coupled-electron pair approximation (CEPA-2) method \cite{starck1993} included in the MOLPRO package \cite{molpro}. Jacobi coordinates have been used to build a grid of 3024 geometries: $r\in [0.8;2.4]$ is the distance between a give pair of nuclei, $R\in [1.9995;56.4227]$ is the distance from the center of mass of the pair to the third nucleus, and $\gamma\in[0;2\pi]$ is the angle between vectors $\overrightarrow{R}$ and $\overrightarrow{r}$. The ground state PES is generated at arbitrary points using an interpolation routine applied to the \textit{ab initio} calculation and an analytical extrapolation at short and large distances. The asymptotic energy of this PES at infinite separation between H$_2(r_e=1.403 \textrm{a.u.})$ and H$^-$ is $D_{as}=-1.701683$a.u. with respect to the origin taken when all particles are infinitely separated. The PES has a van der Waals minimum for the collinear geometry at $r=1.421$~a.u. and $R=6.069$~a.u. with a well of depth 401~cm$^{-1}$ (with respect to $D_{as}$). There is also a barrier for nuclei exchange with a minimum height of $E_h=$3641~cm$^{-1}$ (above $D_{as}$) for collinear geometry at $r_{\text{H}-\text{H}}=r_{\text{H}-\text{H}^-}=1.996$~a.u..

The nuclear dynamics is solved in the centre-of-mass frame assuming $J=0$. We have used equal masses $m_H=1837.3621$~a.u. for all three atoms in the H$_{3}^-$ molecule, \textit{i.e.} the sum of the hydrogen mass and one third of electron mass as required by the hyperspherical approach. In the hyperspherical treatment $\rho$ varies in the interval $[0,\rho_{max})=120.42$. The vibrational and resonant states are characterized according to one of the irreducible representations $\Gamma=$ $A'_1$, $A''_1$, $A'_2$, $A''_2$, $E'$ or $E''$ of the molecular symmetry group $D_{3h}$. In case of $J=0$, the calculation can be performed in $C_{3v}$ symmetry subgroup for H$_3^-$. The hyperangle $\theta$ varies in the interval $[0,\pi/2]$ while the range for $\phi$ can be restricted for three identical particles to $[-\pi/2,-\pi/6]$ for wave function of the $A'_1$ or $A'_2$ irreducible representations (\textit{irreps} in the following) of their molecular symmetry group ($C_{3v}$), and to the interval $[-\pi/2,\pi/2]$ for wave functions of the $E'$ irrep.

All results in the following tables correspond to a total angular momentum $J=0$ with its projection $\Omega=0$ along the axis of $R$. In addition we label the bound and resonant levels with the \textit{irrep} $\Gamma$, and the approximate quantum numbers of the internal state of the dimer $v_d$ and $j_d$ when it is infinitely separated from the negative ion (\textit{i.e.} the dissociation limit they are related to). When a series of levels converges toward a given limit H$_2(v_{d},j_d)$+H$^-$, they are numbered with the integer $v_t \ge 0$ denoting approximately the vibrational motion of the van der Waals complex in the $R$ Jacobi coordinate. For instance the bound levels of $A'_1$ symmetry are labeled with $v_d=0$, $j_d=0$ and $v_t$, while those of $A'_2$ symmetry are labeled with $v_d=0$, $j_d=1$ and $v_t$. The binding energy of the H$_3^-$ bound levels have already been presented in papers I and II, and were found in good agreement with the previous determination of Ref. \cite{starck1993}.

\begin{figure}[t]
\begin{center}
\begin{subfigure}{.5\textwidth}
ù  \centering
  \includegraphics[width=\linewidth]{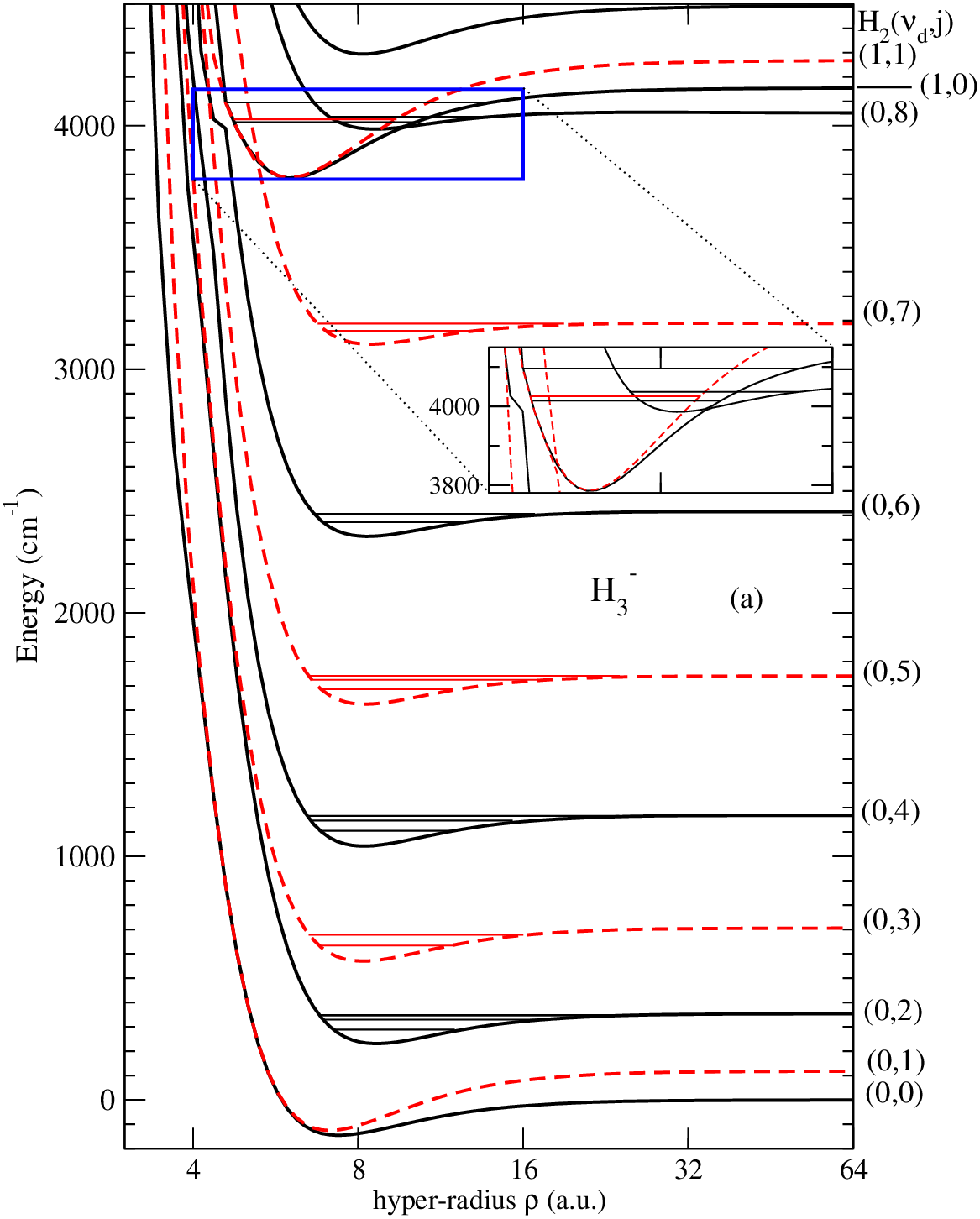}
%  \caption{A subfigure}
% \label{fig:H3-res}
\end{subfigure}%
\begin{subfigure}{.5\textwidth}
%  \centering
  \includegraphics[width=\linewidth]{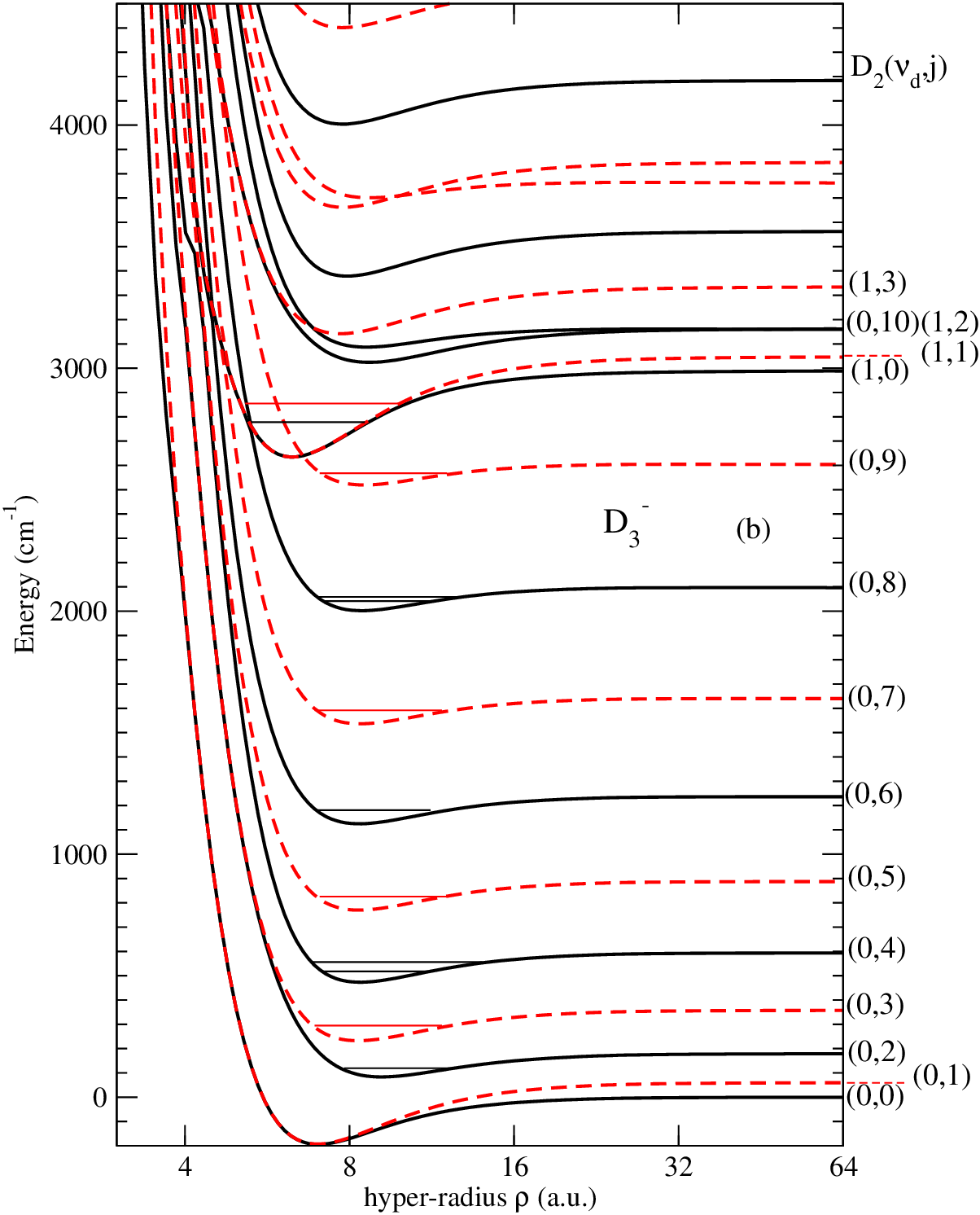}
%  \caption{A subfigure}
% \label{fig:D3-res}
\end{subfigure}%
\end{center}
\caption{Hyperspherical adiabatic potentials of the (a) H$_3^-$ and (b) D$_3^-$ electronic ground state computed in the present work. Dissociation limits are labeled with the vibrational ($v_d$) and rotational ($j_d$) quantum number of H$_2$ (D$_2$). Full black lines: $A'_1$ symmetry; dashed red lines: $A'_2$ symmetry. The curves for $E'$ symmetry are not reported as they would be superimposed on the $A'_1$ and $A'_2$ curves. The actual positions of some of the prereacting resonances displayed in Tables \ref{tab:H3-A1A2} are represented by horizontal lines. Note the logarithmic scale for the hyper-radius axis.}
\label{fig:H3-res}
\end{figure}

The other discrete levels of the ground state PES are prereacting resonances as they are located above the energy $E^d_{00}$ of the H$_2(v_{d}=0,j=0)$+H$^-$ (for $A'_1$ symmetry) or the energy $E^d_{01}$ of the H$_2(v_{d}=0,j=1)$+H$^-$ (for $A'_2$ symmetry) dissociation limits. The corresponding hyperspherical adiabatic potentials are displayed in Fig.\ref{fig:H3-res} and their positions and their widths are given in Table \ref{tab:H3-A1A2}. Potentials for the $E'$ symmetry are not reported in the Figure or in the Table as they are undistinguishable from those of $A'_1$ and $A'_2$ symmetries in the energy range below the top of the barrier (\textit{i.e.} without proton exchange). Note that the minimum height $E_h$ of the barrier above is given with respect to the H$_2(r_e)$+H$^-$ dissociation limit with energy $D_{as}$, while the lowest dissociation limit H$_2(v_{d}=0,j_d=0)$+H$^-$ is located 2173.5~cm$^{-1}$ above $D_{as}$. Therefore all resonances with energies larger than 1468~cm$^{-1}$ above H$_2(v_{d}=0,j_d=0)$+H$^-$ may in principle undergo tunneling above the barrier for a given geometry. However the average over the hyperspherical angles yields a barrier with a top located about 3600~cm$^{-1}$ above $E^d_{00}$. In this energy range we indeed see avoided crossings between the short-range repulsive part of the hysperspherical adiabatic curves correlated to different nuclei arrangements which are the sign of efficient tunneling.

All the identified resonances in a range of about 4000~cm$^{-1}$ above H$_2(v_{d}=0,j_d=0)$+H$^-$ exhibit lifetimes at least in the nanosecond range, so that they may be detectable during a collision provided that the collision energy is well-defined at the millikelvin level. This could be reached for H$^-$ kinetic energy in ion traps, but probably much harder for the neutral kinetic energy. We note that the lifetimes increase with increasing values of the asymptotic dimer rotational quantum number which reflects that the energy gap with open decay channels increases as well. There is no clear evidence for tunneling as the lifetimes do not become shorter when their energy reaches the minimum height of the barrier. When two thresholds are close in energy one could expect the manifestation of Feshbach resonances. An example is displayed in the enlarged view of Fig.\ref{fig:H3-res}(a) with the limits H$_2(v_{d}=1,j_d=0)$+H$^-$ and H$_2(v_{d}=0,j_d=8)$+H$^-$. However the large computed lifetimes suggest that the interaction is probably small which is not surprising given the huge difference in the rotation quantum number of the dimer. Due to the expected large permanent dipole moment, these resonances could be excited once H$_3^-$ is created and trapped, so that they could be excited although their energy is not easily accessible with currently available lasers. Selection rules are the same than for the radiative association process discussed in paper II. For comparison, we also report in Fig.\ref{fig:H3-res}(b) the curves for D$_3^-$ showing the expected influence of the mass on the density of resonances over the same energy range. We assumed $m_D= 3670.8162$~a.u. for the deuterium atom, \textit{i.e.} the sum of the deuterium mass and one third of electron mass.

\begin{table}
\begin{tabular}{|c|c|c|c||c|c|c|c|} \hline
$J$,$\Omega$,$j_d$,$v_{\text{t}}$,$v_{\text{d}}$,$\Gamma$&$E_r$&$\frac{\Upsilon}{2}$&$\tau$ &
$J$,$\Omega$,$j_d$,$v_{\text{t}}$,$v_{\text{d}}$,$\Gamma$&$E_r$&$\frac{\Upsilon}{2}$&$\tau$  \\ \hline
0,0,2,0,0,$A_1'$ & 289.9 &0.0194  & 0.136&0,0,3,0,0,$A_2'$ & 634.2 &0.00004 &64 \\
0,0,2,1,0,$A_1'$ & 330.0 &0.0187  & 0.141&0,0,3,1,0,$A_2'$ & 678.7 &0.000048&54 \\
0,0,2,2,0,$A_1'$ & 348.7 &0.0097  & 0.273&0,0,5,0,0,$A_2'$ & 1686.3&0.00045 & 5.8\\
0,0,4,0,0,$A_1'$ & 1105.2&0.0152  & 0.174&0,0,5,1,0,$A_2'$ & 1725.3&0.00045 & 5.8\\
0,0,4,1,0,$A_1'$ & 1147.4&0.0156  & 0.169&0,0,5,2,0,$A_2'$ & 1740.7&0.0001   &25\\
0,0,4,2,0,$A_1'$ & 1166.0&0.007   & 0.365&0,0,7,0,0,$A_2'$ & 3158.7&0.0000261&101\\
0,0,6,0,0,$A_1'$ & 2372.6&0.00016 & 16   &0,0,7,1,0,$A_2'$ & 3190.5&0.000018& 143\\
0,0,6,1,0,$A_1'$ & 2407.3&0.00014 & 18   &0,0,1,0,1,$A_2'$ & 4026.5&0.0172 &0.154 \\
0,0,8,0,0,$A_1'$ & 4037.8&0.00001 & 298  &0,0,1,1,1,$A_2'$ & 4134.3&0.029  &0.090 \\
0,0,0,0,1,$A_1'$ & 4015.9&0.01    & 0.25 &0,0,1,2,1,$A_2'$ & 4203.1&0.0296 &0.089 \\
0,0,0,1,1,$A_1'$ & 4096.7&0.01    & 0.25 &0,0,1,3,1,$A_2'$ & 4245.4&0.020  &0.12 \\
0,0,0,2,1,$A_1'$ & 4137.4&0.01    & 0.25 &0,0,1,4,1,$A_2'$ & 4262.2&0.010  &0.25\\
0,0,0,3,1,$A_1'$ & 4153.5&0.005   & 0.53 &&&&\\
0,0,2,0,1,$A_1'$ & 4416.2&0.0008  & 3.24 &&&&\\
0,0,2,1,1,$A_1'$ & 4463.4&0.001   & 2.5  &&&&\\
0,0,2,2,1,$A_1'$ & 4484.6&0.001   & 2.5  &&&&\\ \hline
\end{tabular}
\caption{Energy position $E_r$ (in cm$^{-1}$) (with respect to the energy $E^d_{00}$), half-width $\frac{\Upsilon}{2}$ and corresponding lifetime $\tau$ (in ns) of the prereacting resonances of $A'_1$ and $A'_2$ symmetry of the H$_3^-$ electronic ground state. The lowest dissociation limit $E^d_{00}$ is located at 2173.5~cm$^{-1}$ above the energy $D_{as}$ at infinite separation of H$_2(r=r_e=1.403)$+H$^-$ (see paper I). The number of digits for the widths is representative of the achieved convergence when varying the CAP parameters. Levels of $E'$ symmetry are not reported as they are degenerate with those of $A'_1$ and $A'_2$ symmetries in the present energy range.}
\label{tab:H3-A1A2}
\end{table}

\section{Vibrational and resonant states of H$_{2}$D$^-$ and D$_{2}$H$^-$}
\label{sec:hetero}

\begin{figure}[t]
\begin{center}
\begin{subfigure}{.5\textwidth}
%  \centering
  \includegraphics[width=\linewidth]{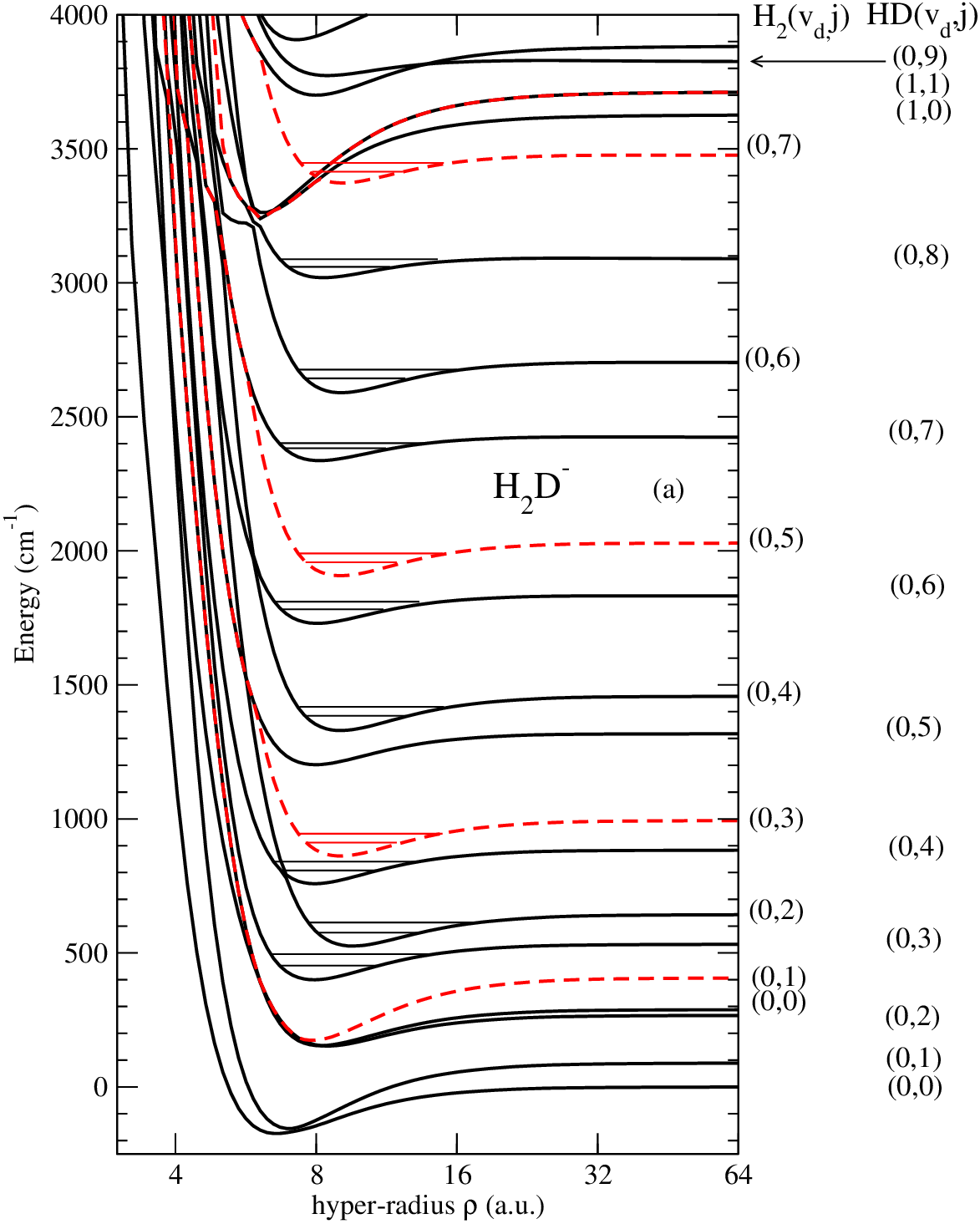}
%  \caption{A subfigure}
% \label{fig:H3-res}
\end{subfigure}%
\begin{subfigure}{.5\textwidth}
%  \centering
  \includegraphics[width=\linewidth]{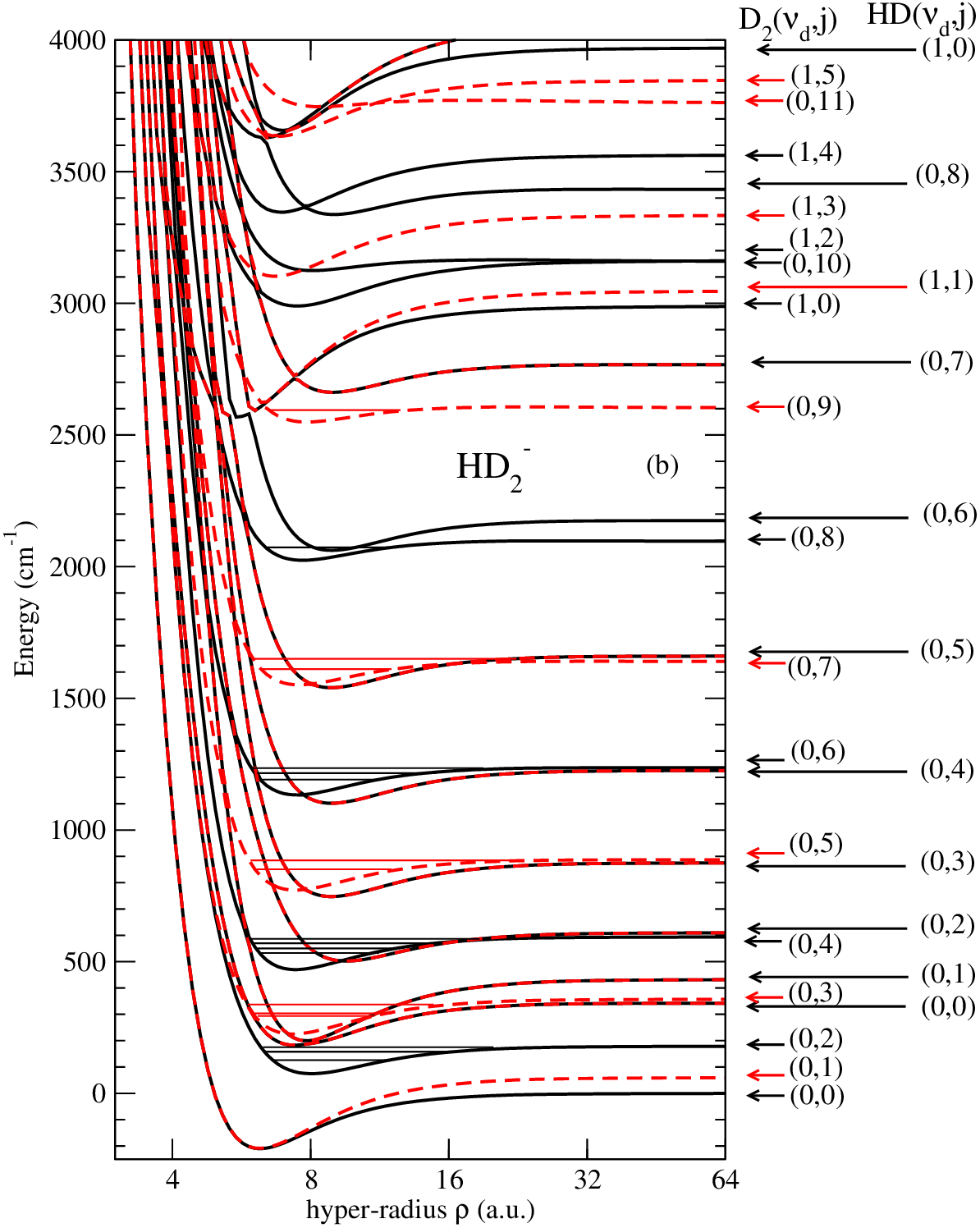}
%  \caption{A subfigure}
% \label{fig:D3-res}
\end{subfigure}%
\end{center}
\caption{Hyperspherical adiabatic potentials of the (a) H$_2$D$^-$ and (b) D$_2$H$^-$ electronic ground state computed in the present work. Dissociation limits are labeled with the vibrational ($v_d$) and rotational ($j_d$) quantum number of the relevant dimer, \textit{i.e.} (a) H$_2$ or HD, and (b) D$_2$ or HD. Full black lines (dashed red lines) refer to the $A$ ($B$) symmetry for states correlated to a dissociation limit involving H$_2$ (D$_2$). Curves related to the $A'$ symmetry (correlated to a limit involving HD) are drawn in full black line and can be distinguished from the previous ones with the arrows on the right side of each panel. The actual positions of some of the prereacting resonances displayed in Tables \ref{tab:H2D-resonances-A'},\ref{tab:H2D-resonances-AB},\ref{tab:D2H-resonances-AB},\ref{tab:D2H-resonances-A'} are represented by horizontal lines. Note the logarithmic scale for the hyper-radius axis.}
\label{fig:H2D-D2H-res}
\end{figure}

The study of the structure of the heteronuclear isotopologues is relevant for experiments. Indeed one can rely on the detection of products of different masses to investigate the dynamics of the trimer. With the same PES the spectroscopic properties of the isotopologues H$_{2}$D$^-$ and D$_{2}$H$^-$ can be extracted by a simple mass scaling using the values of the preceding section to fix reduced masses. In the case of $J=0$, the calculation are performed in $C_{2v}$ symmetry group for both molecules, with $\theta$ in the interval $[0,\pi/2]$ and $\phi$ in the interval $[-\pi/2,\pi/2]$. The upper bound of the hyper-radius grid is fixed at 106.9~a.u. and 92.86~a.u. for H$_2$D$^-$, and D$_2$H$^-$, respectively. The resulting hyperspherical adiabatic curves are displayed in Figure \ref{fig:H2D-D2H-res}. One can immediately see that two kinds of  dissociation limits corresponding to different fragments alternate in energy, making explicit the possibility for tunneling and nucleus exchange reaction. When the fragment dimer is homonuclear the corresponding states are characterized by the $A$ or $B$ \textit{irreps}. On the other hand when HD is the dimer fragment this symmetry cannot be defined anymore. In other words they are represented by a linear combination of states of $A$ and $B$ symmetry yielding a state identified with the label $A'$ of the $Cs$ symmetry group.

We first report in Table \ref{tab:D2H-bound} the energies of the bound levels located below the lowest asymptote in each symmetry, thus extending the results of paper I. As expected from that paper, the results are in good agreement with the previous determination of Ref.\cite{starck1993} for the bound levels of D$_2$H$^-$ located below the the D$_2(0,0)$+H$^-$, D$_2(0,1)$+H$^-$, HD$(0,0)$+H$^-$ limits for the $A$, $B$, and $A'$ symmetries, respectively. We determined the position of a few more levels in the $B$ symmetry compared to Ref.\cite{starck1993}. In principle the latter levels of $A'$ symmetry should be considered as resonances, but we numerically checked that their lifetime is almost infinite. This result expresses that they cannot undergo tunneling through the potential barrier toward the lowest D$_2(v_d=0,j_d=0,1,2)$+H$^-$ outgoing channels. We note also that the uppermost weakly bound levels labeled (0,0,1,5,0,$B$) and (0,0,0,4,0,$A'$) are actually found with a finite lifetime larger than 1000~ns which reflects that their non-adiabatic coupling could  induce their dissociation.

For completeness we display in Table \ref{tab:H2D-bound} the energies of the H$_2$D$^-$ bound levels of symmetry $A'$ located below the lowest HD$(0,0)$+H$^-$ limit, which were previously unknown. Just like the other isotopologue, we checked numerically that the H$_2$D$^-$ levels located below the H$_2(0,0)$+D$^-$ and H$_2(0,1)$+D$^-$ (of $A$ and $B$ symmetry, respectively) can be considered as genuine bound levels with negligible tunneling.

In the last series of Tables \ref{tab:H2D-resonances-A'},\ref{tab:H2D-resonances-AB},\ref{tab:D2H-resonances-AB},\ref{tab:D2H-resonances-A'} we display the energies and widths of H$_2$D$^-$ and D$_2$H$^-$ within the same energy range than for H$_3^-$. A few general trends can be observed. The resonances of $A'$ symmetry in H$_2$D$^-$ and D$_2$H$^-$ are quite short-lived due to the fact that more dissociation channels are open compared to the other cases. For $A$ and $B$ symmetries, resonances in D$_2$H$^-$ are generally larger than those in H$_2$D$^-$, reflecting the higher mass of the system leading to smaller energy gaps between adiabatic channels. Like in the case of H$_3^-$ above, resonances associated to dissociation limits involving a homonuclear dimer with high $j_d$ value are also found almost stable compared to lower resonances.

\begin{table}[h]
\begin{tabular}{|c|c|c|c|c|} \hline
$J$,$\Omega$,$j_d$,$v_{\text{t}}$,$v_{\text{d}}$,$\Gamma$&\multicolumn{2}{|c|}{$E_0$}&\multicolumn{2}{|c|}{$E_b^0(E'_b)$}\\
D$_2$H$^-$&This work&\cite{starck1993}&This work&\cite{starck1993} \\ \hline
0,0,0,0,0,$A$ & 1429.1& 1430.9 &-113.1&-107.3 \\
0,0,0,1,0,$A$ & 1493.4& 1492.5 &-48.8 &-45.7  \\
0,0,0,2,0,$A$ & 1526.5& 1526.3 &-15.7 &-11.9\\
0,0,0,3,0,$A$ & 1540.0& 1536.7 &-2.2 &-1.5\\  \hline
0,0,1,0,0,$B$ & 1433.5  & 1436.5 &-108.7 (-168.4)&-101.7 (n/a) \\
0,0,1,1,0,$B$ & 1509.4  & 1509.6 & -32.8(-92.0)&-28.6 (n/a) \\
0,0,1,2,0,$B$ & 1557.7  &        &15.5 (-44.2)&  \\
0,0,1,3,0,$B$ & 1584.6  &        &42.4 (-17.3)&  \\
0,0,1,4,0,$B$ & 1597.6  &        &55.4 (-4.3 )& \\
0,0,1,5,0,$B$ & 1601.9  &        &59.7 (-0.05)&  \\  \hline
0,0,0,0,0,$A'$ & 1788.1  & 1789.5 & 245.8(-97.1)& 251.3(-91.3) \\
0,0,1,0,0,$A'$ & 1804.2  & 1806.2 & 261.9(-81.0) & 268.0(-74.6)\\
0,0,0,1,0,$A'$ & 1837.3  & 1835.1 & 295.0(-47.9)& 296.9(-45.7) \\
0,0,0,2,0,$A'$ & 1862.7  & 1864.3 & 319.8(-23.1) &326.1(-16.5) \\
0,0,1,1,0,$A'$ & 1870.8  & 1867.2 & 327.4(-15.5) & 329.0(-13.6) \\
0,0,0,3,0,$A'$ & 1880.7  & 1877.1 & 337.6(-5.3)  & 338.9(-3.7)  \\
0,0,0,4,0,$A'$ & 1885.2  & 1880.3 & 342.9(-0.06) & 342.1(-0.5)  \\ \hline
\end{tabular}%
\caption{Computed energies (in cm$^{-1}$) for bound levels of the D$_{2}$H$^-$ molecule. The energies $E_0$ are given with respect to $D_{as}$ (see text). The energies $E_b^0$ are given with respect to the dissociation limit D$_2(0,0)$+H$^-$ for all levels. The energies $E'_b$ relative to D$_2(v_d=0,j_d=1)$+H$^-$ for $B$ levels, and relative to the HD$(v_d=0,j_d=1)$+D$^-$ for the $A'$ levels are also given in parenthesis.}
\label{tab:D2H-bound}
\end{table}

\begin{table}[h]
\begin{tabular}{|c|c|c|c|c|} \hline
$J$,$\Omega$,$j_d$,$v_{\text{t}}$,$v_{\text{d}}$,$\Gamma$&$E_0$&$E_b^0(E'_b)$\\
H$_2$D$^-$&\multicolumn{2}{|c|}{This work} \\\hline
0,0,0,0,0,$A'$& 1797.9 & -87.3 \\
0,0,0,1,0,$A'$& 1816.0 & -69.2\\
0,0,0,2,0,$A'$& 1851.2 & -34.0\\
0,0,0,3,0,$A'$& 1875.5 & -9.7\\
0,0,0,4,0,$A'$& 1884.5 & -0.7\\ \hline
0,0,0,0,0,$A$ & 2095.9 &210.7 (-77.7) \\
0,0,0,1,0,$A$ & 2138.2 &253.0 (-35.4) \\
0,0,0,2,0,$A$ & 2161.4 &276.2 (-12.2)\\
0,0,0,3,0,$A$ & 2171.8 &286.6 (-1.8)  \\  \hline
0,0,1,0,0,$B$ & 2129.5  &244.3 (-162.3) \\
0,0,1,1,0,$B$ & 2196.2  &311.0 (-95.6) \\
0,0,1,2,0,$B$ & 2241.3  &356.1 (-50.5)  \\
0,0,1,3,0,$B$ & 2268.8  &383.6 (-23.0) \\
0,0,1,4,0,$B$ & 2283.8  &398.6 (-8.0) \\
0,0,1,5,0,$B$ & 2290.6  &405.4 (-1.2) \\ \hline
\end{tabular}%
\caption{Computed energies (in cm$^{-1}$) for bound levels of the H$_{2}$D$^-$ molecule. The energies $E_0$ are given with respect to $D_{as}$ (see text). The energies $E_b^0$ are given with respect to the dissociation limit HD$(v_d=0,j_d=0)$+H$^-$ for all levels. The energies $E'_b$ relative to H$_2(v_d=0,j_d=0)$+H$^-$ for $A$ levels, and to H$_2(v_d=0,j_d=1)$+H$^-$ for $B$ levels are also given in parenthesis.}
\label{tab:H2D-bound}
\end{table}

\begin{table}[h]
\begin{tabular}{|c|c|c|c|c|} \hline
$J$,$\Omega$,$j_d$,$v_{\text{t}}$,$v_{\text{d}}$,$\Gamma$&$E_r$&$E_r^0$&$\frac{\Upsilon}{2}$&$\tau$ \\ \hline
0,0,2,0,0,$A'$ & 2092.0 & 206.8& 0.0189 & 0.140 \\
0,0,2,1,0,$A'$ & 2129.1 & 243.9& 0.0171 & 0.155 \\
0,0,2,2,0,$A'$ & 2148.2 & 263.0& 0.0053 & 0.501\\
0,0,3,0,0,$A'$ & 2389.2 & 504.0& 0.2307 & 0.0115\\
0,0,3,1,0,$A'$ & 2413.0 & 527.8& 0.1223 & 0.0217\\
0,0,4,0,0,$A'$ & 2704.5 & 819.3& 0.0310  & 0.0856\\
0,0,4,1,0,$A'$ & 2745.7 & 860.5& 0.0362  & 0.0733\\
0,0,4,2,0,$A'$ & 2766.6 & 881.4& 0.0091  & 0.291\\
0,0,5,0,0,$A'$ & 3147.2 & 1262.0& 0.0166 & 0.160\\
0,0,5,1,0,$A'$ & 3185.9 & 1300.7& 0.0165 & 0.160\\
0,0,6,0,0,$A'$ & 3672.4 & 1787.2& 0.0062 &  0.43\\
0,0,6,1,0,$A'$ & 3707.7 & 1822.5& 0.0050 &  0.53\\
0,0,7,0,0,$A'$ & 4276.5 & 2391.3& 0.0024 & 1.1 \\
0,0,7,1,0,$A'$ & 4307.0 & 2421.8& 0.0023 &  1.1\\
0,0,8,10,0,$A'$& 4955.5 & 3070.3& 0.0007 &  3.8\\  \hline
\end{tabular}
\caption{Computed energies (in cm$^{-1}$), half-widths $\frac{\Upsilon}{2}$ (in cm$^{-1}$) and corresponding lifetimes $\tau$ (in ns) for prereacting resonances in the H$_{2}$D$^-$ molecule. The energies $E_r$ are given with respect to $D_{as}$. The energies $E_r^0$ are given with respect to the dissociation limit HD$(v_d=0,j_d=0)$+H$^-$ located at 1885.2~cm$^{-1}$ above $D_{as}$. }
\label{tab:H2D-resonances-A'}
\end{table}

\begin{table}[h]
\begin{tabular}{|c|c|c|c|c|} \hline
$J$,$\Omega$,$j_d$,$v_{\text{t}}$,$v_{\text{d}}$,$\Gamma$&$E_r$&$E_r^0(E'_r)$&$\frac{\Upsilon}{2}$&$\tau$ \\ \hline
0,0,2,0,0,$A$ & 2455.9 &570.7(282.3)  & 0.0029& 0.91 \\
0,0,2,1,0,$A$ & 2494.7 &609.5(321.1)  & 0.0045& 0.59\\
0,0,2,2,0,$A$ & 2517.1 &631.9(343.5)  & 0.0018&1.47\\
0,0,2,3,0,$A$ & 2525.9 &640.7(352.3)  & 0.0012&2.2\\
0,0,4,0,0,$A$ & 3264.5 &1379.3(1090.9)& 0.0039&0.68\\
0,0,4,1,0,$A$ & 3306.7 &1421.5(1133.1)& 0.0098 &0.27\\
0,0,4,2,0,$A$ & 3332.8 &1447.6(1159.2)& 0.0051 &0.52\\
0,0,4,3,0,$A$ & 3341.1 &1455.9(1167.5)& 0.0035 &0.76\\
0,0,6,0,0,$A$ & 4523.7 &2638.5(2350.1)& 0.0000264&1005\\
0,0,6,1,0,$A$ & 4563.1 &2677.9(2389.5)& 0.0000267&994\\
0,0,6,2,0,$A$ & 4584.7 &2699.5(2411.1)& 0.0000064&4147\\
0,0,3,0,0,$B$ & 2796.7  &911.5(504.9) & 0.0000051&5200\\
0,0,3,1,0,$B$ & 2839.6  &954.4(547.8) & 0.0000117&226\\
0,0,3,2,0,$B$ & 2867.0  &981.8 (575.2) & 0.0000036&737\\
0,0,3,3,0,$B$ & 2876.1  &990.9(584.3) & 0.0000058&457\\
0,0,5,0,0,$B$ & 3841.6  &1656.4(1549.8) & 0.0000615 &43.1\\
0,0,5,1,0,$B$ & 3882.7  &1997.5(1590.9)& 0.0002565 &10.34\\
0,0,5,2,0,$B$ & 3906.9  &2021.7(1615.1)&  0.0000804&33.0\\
0,0,5,3,0,$B$ & 3914.1  &2028.9(1622.3)& 0.0000382 &69.5\\
0,0,7,0,0,$B$ & 5304.9  &3013.1(3419.7) &  0.0000010 &2654\\
0,0,7,1,0,$B$ & 5342.3  &3457.1(3050.5) & 0.0000054&491\\ \hline
\end{tabular}
\caption{Same as Table \ref{tab:H2D-resonances-A'} for resonances of $A$ and $B$ symmetries. We added in parenthesis the energy $E'_r$ with respect to the dissociation limit H$_2(v_d=0,j_d=0)$+D$^-$ (for $A$ levels) and H$_2(v_d=0,j_d=1)$+D$^-$ (for $B$ levels) located respectively at 2173.6~cm$^{-1}$ and 2291.8~cm$^{-1}$ above $D_{as}$. }
\label{tab:H2D-resonances-AB}
\end{table}

\begin{table}[h]
\begin{tabular}{|c|c|c|c|c|c|c|} \hline
$J$,$\Omega$,$j_d$,$v_{\text{t}}$,$v_{\text{d}}$,$\Gamma$&$E_r$&$E_r^0$&$\frac{\Upsilon}{2}$&$\tau$ \\ \hline
0,0,2,0,0,$A$ & 1669.1 & 126.9  &0.0045& 0.59\\
0,0,2,1,0,$A$ & 1700.7 & 158.5  &0.0061& 0.43\\
0,0,2,2,0,$A$ & 1718.0 & 175.8  &0.0036& 0.73\\
0,0,4,0,0,$A$ & 2075.0 & 532.8  &0.1868& 0.0142\\
0,0,4,1,0,$A$ & 2113.2 & 571.0  &0.1860& 0.0142\\
0,0,4,2,0,$A$ & 2133.5 & 591.3  &0.1008& 0.0263\\
0,0,4,3,0,$A$ & 2137.0 & 594.8  &0.0137& 0.194\\
0,0,6,0,0,$A$ & 2735.0 & 1192.8 &0.0051& 0.52  \\
0,0,6,1,0,$A$ & 2768.1 & 1225.9 &0.0037& 0.72 \\
0,0,6,2,0,$A$ & 2782.2 & 1240.0 &0.0028& 0.95 \\
0,0,8,0,0,$A$ & 3619.3 & 2077.1 &0.0003& 8.8\\
0,0,8,0,0,$A$ & 3643.0 & 2100.8 &0.0001& 26\\
0,0,3,0,0,$B$ & 1831.0 & 288.8(229.1)& 0.000504 &5.26 \\
0,0,3,1,0,$B$ & 1871.0 & 328.8(269.1)& 0.000524  & 5.06 \\
0,0,3,2,0,$B$ & 1893.5 & 251.3 (291.6)&0.000496 & 5.35\\
0,0,3,3,0,$B$ & 1899.5 & 357.3 (297.6)&0.000078  & 34\\
0,0,5,0,0,$B$ & 2375.4 & 833.2 (773.5)&0.0052 & 0.51 \\
0,0,5,1,0,$B$ & 2411.6 & 1869.4 (809.7)&0.0060 & 0.44 \\
0,0,5,2,0,$B$ & 2429.6 & 787.4 (827.7)& 0.0019 & 1.4\\
0,0,7,0,0,$B$ & 3150.2 & 1608.0 (1548.3)&0.0013 &2.04\\
0,0,7,1,0,$B$ & 3178.5 & 1636.3 (1576.6)& 0.000636& 4.17\\
0,0,9,0,0,$B$ & 4139.8 & 2597.6 (2537.9)& 0.00010 &26\\  \hline
\end{tabular}
\caption{Computed energies (in cm$^{-1}$), half-widths $\frac{\Upsilon}{2}$ (in cm$^{-1}$) and corresponding lifetimes $\tau$ (in ns) for prereacting resonances in the D$_{2}$H$^-$ molecule. The energies $E_r$ are given with respect to $D_{as}$. The energies $E_r^0$ are given with respect to the dissociation limit D$_2(v_d=0,j_d=0)$+H$^-$ located at 1542.2~cm$^{-1}$ above $D_{as}$. For resonances of $B$ symmetry we added in parenthesis the energy $E'_r$ with respect to the dissociation limit D$_2(v_d=0,j_d=1)$+H$^-$  located at 1601.9~cm$^{-1}$ above $D_{as}$.}
\label{tab:D2H-resonances-AB}
\end{table}

\begin{table}[h]
\begin{tabular}{|c|c|c|c|c|c|c|} \hline
$J$,$\Omega$,$j_d$,$v_{\text{t}}$,$v_{\text{d}}$,$\Gamma$&$E_r$&$E_r^0$&$\frac{\Upsilon}{2}$&$\tau$ \\ \hline
0,0,2,0,0,$A'$ & 2084.4 &542.2 (192.2)& 0.0019  & 1.40 \\
0,0,2,1,0,$A'$ & 2118.2 &576.0 (233.0)&  0.0018  & 1.47 \\
0,0,2,2,0,$A'$ & 2137.8 &595.6 (252.6)&  0.0016  &1.66  \\
0,0,2,3,0,$A'$ & 2148.4 & 606.2 (263.2)&  0.0013 & 2.04 \\
0,0,3,0,0,$A'$ & 2335.5 & 793.3 (450.3)& 0.1693  & 0.0156 \\
0,0,3,1,0,$A'$ & 2374.7 &832.5 (489.5)& 0.2617  & 0.0101 \\
0,0,3,2,0,$A'$ & 2401.2 &859.0 (516.0)& 0.2175  & 0.0122 \\
0,0,3,3,0,$A'$ & 2411.7 &869.5 (526.5)&  0.0969  & 0.0273 \\
0,0,3,4,0,$A'$ & 2417.2 &875.0 (532.0)& 0.0218   & 0.121 \\
0,0,4,0,0,$A'$ & 2689.7 &1147.5 (804.5)& 0.0062  & 0.43\\
0,0,4,1,0,$A'$ & 2728.7 &1186.5 (843.5)&  0.0087 & 0.30 \\
0,0,4,2,0,$A'$ & 2754.5 &1212.3 (869.3)&  0.0079 & 0.33 \\
0,0,4,3,0,$A'$ & 2764.1 &1221.9 (878.9)&  0.0042& 0.63  \\
0,0,5,0,0,$A'$ & 3128.3 &1586.1 (1243.1)&  0.0060 & 0.44 \\
0,0,5,1,0,$A'$ & 3166.4 &1624.2 (1281.2)&  0.0098 & 0.27 \\
0,0,5,2,0,$A'$ & 3191.3 & 1649.1 (1306.1)& 0.0080  & 0.33\\
0,0,5,3,0,$A'$ & 3199.9 & 1657.7 (1314.7)& 0.0042  & 0.63 \\
0,0,6,0,0,$A'$ & 3649.0 &2106.8 (1763.8)&   0.0016 & 1.66 \\
0,0,6,1,0,$A'$ & 3685.9 &2143.7 (1800.7)&   0.0026&  1.02\\
0,0,6,2,0,$A'$ & 3709.2 &2167.0 (1824.0)&   0.0018&  1.47\\
0,0,6,3,0,$A'$ & 3716.7 &2174.5 (1831.5)&   0.0010&  2.6\\
0,0,7,0,0,$A'$ & 4248.3 &2706.1 (2363.1)&   0.00067& 3.96 \\
0,0,7,1,0,$A'$ & 4283.6 &2741.4 (2398.4)&   0.00096&  2.76\\
0,0,7,2,0,$A'$ & 4304.7 &2762.5 (2419.5)&   0.00092&  2.88\\ \hline
\end{tabular}
\caption{Same as Table \ref{tab:H2D-resonances-AB} for resonances of $A'$ symmetry. We added in parenthesis the energy $E'_r$ with respect to the dissociation limit  HD$(v_d=0,j_d=0)$+D$^-$ located at 1885.2~cm$^{-1}$ above $D_{as}$.}
\label{tab:D2H-resonances-A'}
\end{table}

\section{Discussion and perspectives}
\label{sec:conclusion}
In this paper we characterized the complex structure of the bound and resonant levels of H$_3^-$ and its isotopologues H$_2$D$^-$ and D$_2$H$^-$ within an energy range of about 4000~cm$^{-1}$ above the lowest dissociation limit where the relevant diatom is in its rovibrational ground level $v_d=0,j_d=0$. This energy range is mostly located below the top of the potential barrier between the two configurations where one nucleus has been exchanged between the atomic negative ion and the neutral dimer (\textit{i.e.} H$_2$/H$^-$, H$_2$/D$^-$ and D$_2$/H$^-$ respectively). We took advantage of the elegant formalism of the Slow Variable Discretization (SVD) \cite{tolstikhin1996} to implement the calculation of energy levels and widths in the body-fixed frame of the trimer for states with total angular momentum $J=0$ using the hyperspherical coordinates. We extended the preliminary study of Ref.\cite{blandon2007} for three-body resonances to the realistic case of the family of the simplest triatomic negative ions. The present results contribute to their improved knowledge in the perspective of the search for its possible presence in astrophysical environments \cite{ayouz2011}, in the context of the recent discovery of stable molecular anions in the interstellar medium \cite{mccarthy2006,agundez2010}.

The level structure we obtained is conveniently labeled by the dissociation limit of the hyperspherical adiabatic curves in terms of the internal rovibrational state of the fragment dimer. This first expresses that the couplings between various adiabatic channels $N_c$ is generally small, inducing non-radiative lifetimes for the resonance larger than 0.1~ns. For every adiabatic hyperspherical curve, we identified a vibrational structure associated to the relative vibration of the negative ion and of the dimer along the axis connecting their centre-of-mass. This corresponds to the long-range nature of the interaction between the negative ion and the neutral dimer, dominated by the charge-quadrupole interaction varying as $R^{-3}$. The rotational structure of these levels $v_t$, reflecting the mechanical rotation of the negative ion and the dimer, can readily be estimated considering the position of the minima of the hyperspherical adiabatic curves around 8~a.u. for the bound levels, and around 10~a.u. for the resonances. Assuming that $\rho \approx R$ and considering the reduced mass $\mu_{Jacobi}$ of the ion-dimer system, the resulting rotational constant $B_e=1/(2\mu_{Jacobi} R^2)$ is found in the range of 1~cm$^{-1}$ for H$_3^-$ down to 0.4~cm$^{-1}$ for the heavy isotopologues. This value is well below the typical energy spacing between two of such vibrational levels $v_t$. It is worthwhile to recall that imposing $J=0$ in these kinds of calculations involving hyperspherical coordinates imposes that the orbital angular momentum $\vec{\ell}$ between the negative ion and the dimer is equal to the internal angular momentum $\vec{j_d}$ of the dimer. From a collisional point of view, the reported results correspond to increasing partial waves as the dissociation limit is higher in energy. Thus the hyperspherical adiabatic curves actually behave asymptotically as $E_d+\ell(\ell+1)/(2\mu_{Jacobi} R^2)$ (assuming $\rho \approx R$ at large distances) where $E_d$ is the energy of the dimer. The small widths of the resonances suggest that for $J=0$, $\ell \equiv j_d$ is a good quantum number, and would actually determine the rotational structure of these resonances.

Up to now there is no evidence for the observation of such resonances in the scattering experiments between D$_2$ and H$^-$, or H$_2$ and D$^-$ \cite{haufler1997}. First the energy resolution was such that only patterns associated to the internal state of the fragment dimer were characterized, either its vibrational state \cite{haufler1997} or its rotational state \cite{muller1996}. Furthermore these experiments have been performed for collision energies higher than the predicted potential barrier so that reactive processes are likely to dominate the dynamics.

In order to check the present results, high-resolution scattering experiments are desirable, which could be achieved soon using multipole traps for cold ions. Moreover the present calculations should be extended to non-zero total angular momentum (see for instance Ref.\cite{quemener2005}) in order to compute total cross sections including all relevant partial waves, to predict the actual contribution of such resonances.

\begin{acknowledgments}
We thank Pr Viatcheslav Kokoouline for providing the SVD code in hyperspherical coordinates for calculation of vibrational resonances. Stimulating discussions with Prof. Roland Wester and Dr Maurice Raoult are gratefully acknowledged. The study was partially supported by the \textit{R\'eseau th\'ematique de recherches avanc\'ees "Triangle de la Physique"}, contract QCCM-2008-007T, the National Science Foundation with grant No PHY-08-55622. We used resources of the National Energy Research Scientific Computing Center, which is supported by the Office of Science of the U.S. Department of Energy under contract No. DE-AC02-05CH11231.
\end{acknowledgments}

\newpage
%\bibliography{../DATABASE/bibliocold}
%Merlin.mbs v4.21 2009-07-09.
%

\end{document}